\documentclass[smallextended]{svjour3}      

\smartqed  
\usepackage{graphicx}
\usepackage{hyperref}
\usepackage[colorinlistoftodos]{todonotes}
\usepackage{color}
\usepackage{amsmath}
\usepackage{amsfonts}
\usepackage{amssymb}

\begin{document}

\title{Emergence of time 
}

\author{George F R Ellis 
and Barbara Drossel   
}

\institute{George F R Ellis \at
              Mathematics Department, University of Cape Town \\
              Rondebosch, Cape Town 7701, South Africa\\
              \email{george.ellis@uct.ac.za} 
              \\    \\     %
	Barbara Drossel \at Institute of Condensed Matter Physics\\
	Technische Universit\"at Darmstadt\\
	Hochschulstr. 6, 64289 Darmstadt, Germany\\
\email{drossel@fkp.tu-darmstadt.de}           
}
\date{Received: date / Accepted: date}

\maketitle

\begin{abstract}
\textit{Microphysical laws are time reversible, but macrophysics, chemistry and biology are not. This chapter explores how this asymmetry (a classic example of a broken symmetry) arises due to the cosmological context, where a non-local Direction of Time is imposed by the expansion of the universe. This situation is best represented by an Evolving Block Universe, where local arrows of time (thermodynamic, electrodynamic, gravitational, wave,  quantum, biological) emerge in concordance with the Direction of Time because a global Past Condition results in the Second Law of Thermodynamics pointing  to the future. At the quantum level, the indefinite  future  changes  to  the  definite  past  due  to  quantum wave function collapse  events.}

\keywords{Evolving Block Universe \and Arrow of time \and Direction of time \and Wave Function Collapse \and Quantum Gravity}

\end{abstract}

\section{Introduction}
\label{sec:Intro} The nature  of time 
is a contested issue, with some claiming time does not pass for a variety of reasons, and some claiming the opposite (see \cite{Barbour} \cite{SciAm} and \href{https://en.wikipedia.org/wiki/Eternalism_(philosophy_of_time)}{Wikipedia: Eternalism}). This chapter claims that the passage of time is real, and is based at the micro level in quantum wave function collapse whereby the indefinite future becomes the definite past. 

Section \ref{sec:non-UNitary} discusses how time passes: unitary evolution, often used to dispute this, is the exception in the real world.  Section \ref{sec:EBU} explains the associated spacetime view, namely the \textit{Evolving Block Universe} (EBU), and   the emergence of the (global)
\textit{Direction of Time} due to the cosmological context in which we live.   
Section \ref{sec:arrows} discusses the emergence of the various (local) \textit{Arrows of Time}, in agreement with the Direction of Time.  Section \ref{sec:quantum} discusses how this relates to the quantum collapse of the wave function in a semi-classical view (spacetime is classical), while  Section \ref{sec:quanr_grav}
discusses quantum gravity issues arising because quantum spacetime itself is emergent. 
Section \ref{sec:conclude} summarizes the outcomes.

\section{Time passes: unitary is the exception}\label{sec:non-UNitary}
A claim made by a number of writers is that 
time does not pass because physical evolution, whether classical or quantum,  is Hamiltonian and unitary. The situation  in mind 
 is a Hamiltonian  system (classical or quantum) with a time independent potential term:  \begin{equation}\label{eq:Hamiltonian}
 V = V(\textbf{q}) \,\,\Rightarrow \,\,H  = H(\textbf{p},\textbf{q}).
\end{equation}
The equations of motion
\begin{equation}\label{eq:Hamiltonian_motion}
\frac{d\textbf{q}}{d t}  = \frac{\partial H}{\partial \textbf{p}},\,\, \frac{d\textbf{p}}{d t}  = -\frac{\partial H}{\partial \textbf{q}}.
\end{equation}
  imply the dynamics is deterministic and time reversible. Indeed,  initial data $\{\textbf{x}_0,\textbf{p}_0\}$ for such a system given at time $t_0$ uniquely determines the solution 
$\{\textbf{x}(t),\textbf{p}(t)\}$ for all times \cite{Arnold}:
\begin{equation}\label{eq:unitary}
\{\textbf{x}_0,\textbf{p}_0\} \Rightarrow \{\textbf{x}(t),\textbf{p}(t)\}\,\,\forall\,\, t
\end{equation}
 so there is no special meaning to the present time $t_0$ and hence time does not pass. One can eliminate the parameter $t$ from the dynamical relation $\{\textbf{x}(t),\textbf{p}(t)\}$ to get the phase plane relation $\{\textbf{p}(\textbf{x})\}$, and time has disappeared from the solution. 

Now there are several problems with this. It does not do justice to the  micro-macro issue (\S \ref{sec:micro_macro}), it does not take the role of time dependent  constraints into account (\S\ref{sec:_time_constraint}), it ignores system dissipative/open nature due to ongoing coupling to the environment (\S\ref{se: env_coupling}), and  
 it does not take quantum uncertainty and wave function collapse into account 
 discussed in Section \ref{sec:quantum}). Furthermore, a deterministic description of the form \eqref{eq:unitary} presupposes that the initial state possesses infinite precision, i.e., that it contains an infinite amount of information. However, this is impossible in a universe that has a finite density of information \cite{gisin2018indeterminism} and where no physical infinities occur \cite{Ellisetal_infinity}. This issue is particularly relevant for non-integrable systems, for which a limited precision of the initial state does not allow a prediction of the future time evolution beyond a certain time horizon. 

 Because of these various issues, unitary evolution (\ref{eq:unitary}) is the exception rather than the rule in the real world. One has $\{\textbf{p}(\textbf{x},t)\}$ rather than $\{\textbf{p}(\textbf{x})\}$. This is true both as regards the parts, and the whole (Section \ref{sec:parts_whole}).
 
\subsection{Macro-micro issues}\label{sec:micro_macro}
Significant issues arise in terms of the way time works at different levels in the hierarchy of emergence and complexity \cite{Ell16} on both the natural science and biological science sides. The key point is that things happen irreversibly and time asymmetrically at the macro scale, as emphasized strongly by Arthur Eddington \cite{Eddington}, even though our description of the foundational dynamics  at the microscale (based in Hamiltonians) is reversible and time symmetric.
Things are not unitary at the macro scale, except in very rare circumstances for restricted times, such as the motion of the planets and comets round the Sun.

 The most famous macroscopic example is the thermodynamic arrow of time, as expressed in the Second Law of Thermodynamics: 
 \begin{equation}\label{eq:2nd_Law}
 dS/dt \geq 0
 \end{equation} 
 where $S$ is the entropy of an isolated system. This is very important in physical chemistry, biochemistry, biology in general, and  social life (where it underlies the basic resource issues facing us), as well as in astrophysics and cosmology. Irreversible processes take place during baryosynthesis, at the end of inflation, during nucleosynthesis, decoupling of matter and radiation, and star formation and evolution \cite{Dod03} \cite{PetUza13}. Thus time undoubtedly passes at the macro scale, with information being lost  through dissipative processes (for example, once a pendulum has come to a stop, you can't determine what its past motion was from its present state).  It is fundamentally important that one cannot determine in what direction of time (\ref{eq:2nd_Law}) will hold by coarse graining either classical or quantum micro physics, because of the time symmetry of the relevant equations; rather this is determined in a contextual way  by macroscopic conditions. 
 We return to this fundamental issue (Loschmidt's paradox) in Section \ref{sec:time_Thermod}. 
 
 However equally important are the electrodynamic (\S\ref{sec:time_Electrodynamics}), wave (\S\ref{sec:arrow-waves})  and quantum (\S\ref{sec:time_quantum}) arrows of time.
The gravitational arrow of time (\S\ref{sec:time_grav}) is important in some astrophysical contexts, but not in the Solar System at present or in daily life (it relates to situations where  gravitational radiation is important).

The thermodynamic, electrodynamic, wave, and quantum arrows of time jointly lead to the biological arrow of time (\S\ref{sec:time_biol}), and presumably the mental arrow of time, which we incontrovertibly all experience in our daily lives; as stated  expressively by Omar Khay\'{y}am:
\begin{quote}
	``\textit{“The Moving Finger writes; and, having writ,\\
		Moves on: nor all thy Piety nor Wit\\
		Shall lure it back to cancel half a Line,\\
		Nor all thy Tears wash out a Word of it.”}
\end{quote}
The position we will take is that the passage of time at the macro scale is irrefutable, and how this relates to 
the microphysics must be taken as a key issue one must explain. Our answer will be that top down causation takes place in the hierarchy of structure and causation \cite{Ell16}, where a time varying cosmology is the context for all structure formation and sets the arrows of time for local processes at both micro and macro levels.
Time dependent constraints (\S\ref{sec:_time_constraint}) and ongoing coupling to the environment (\S\ref{se: env_coupling}) both occur, so  (\ref{eq:unitary}) does not hold.

\paragraph{The Hierarchies of structure and causation}
To discuss this properly, one must clearly distinguish  the different levels of the emergent hierarchy of structure and causation in the natural sciences \cite{Ell16} (left side of Table \ref{Table2}) and biological sciences \cite{Nob12} \cite{Ell16} (right side of Table  \ref{Table2}). Different kinds of causation emerge at each level of these hierarchies, described by different kinds of variables. The higher levels on the natural sciences side are characterised by larger scales and lower interaction energy per volume between the objects that are relevant for the hierarchical level under consideration.
At higher levels processes come into play that cannot be described in lower level terms; formation of a galaxy, or accretion processes onto a black hole, for example.  

 \vspace{0.1in}
\begin{tabular}{|c|c|c|c|}
	\hline \hline
	& Inanimate matter & Living matter \\	
	\hline Level 10 & Cosmology &  Sociology/Economics/Politics \\ 
\hline 
	Level 9	& Astronomy &  Psychology \\ 
\hline 
	Level 8	& Space science &  Physiology \\ 
\hline 
	Level 7	& Geology, Earth Science &  Cell Biology \\ 
	\hline 
	Level 6	& Materials Science & Molecular biology  \\ 
	\hline 
	Level 5	& Physical chemistry & Biochemistry \\ 
	\hline 
	Level 4 & Atomic Physics	& Atomic Physics
	\\ 
	\hline 
	Level 3	& Nuclear Physics & Nuclear Physics  \\ 
	\hline 
	Level 2	&  Particle Physics & Particle Physics  \\ 
	\hline 
	Level 1	& Fundamental Theory & Fundamental Theory \\ 
		\hline \hline
\end{tabular} 
\vspace{0.1in}

\textbf{Table 2.1}:\label{Table2} \textit{The emergent hierarchy of structure and causation for inanimate matter (left) and life (right) as characterised by academic discipline}  \cite{Ell16}.\\

On the life sciences side, quite different kinds of causation come into play at higher levels \cite{CamRee05}  related to biochemical and molecular  biology (Level 6), cellular processes (Level 7), physiological processes (Level 8), mental and
psychological processes (Level 9), and social processes (Level 10), each of which
is clearly causally effective; for example mental processes lead to the construction of aircraft and digital computers \cite{Ell16}, which would not otherwise exist \cite{Ell05}. 
There are similar hierarchies in the case of complex artificial systems \cite{SimonHA} such as watches, automobiles, aircraft, cities, and digital computers, because all complex systems are modular hierarchical structures. The  case of digital computers is discussed in detail in \cite{Tannen} and \cite{Ellis_Drossel}.

\begin{figure}[h]
	\centering
	\includegraphics[width=0.8\linewidth]{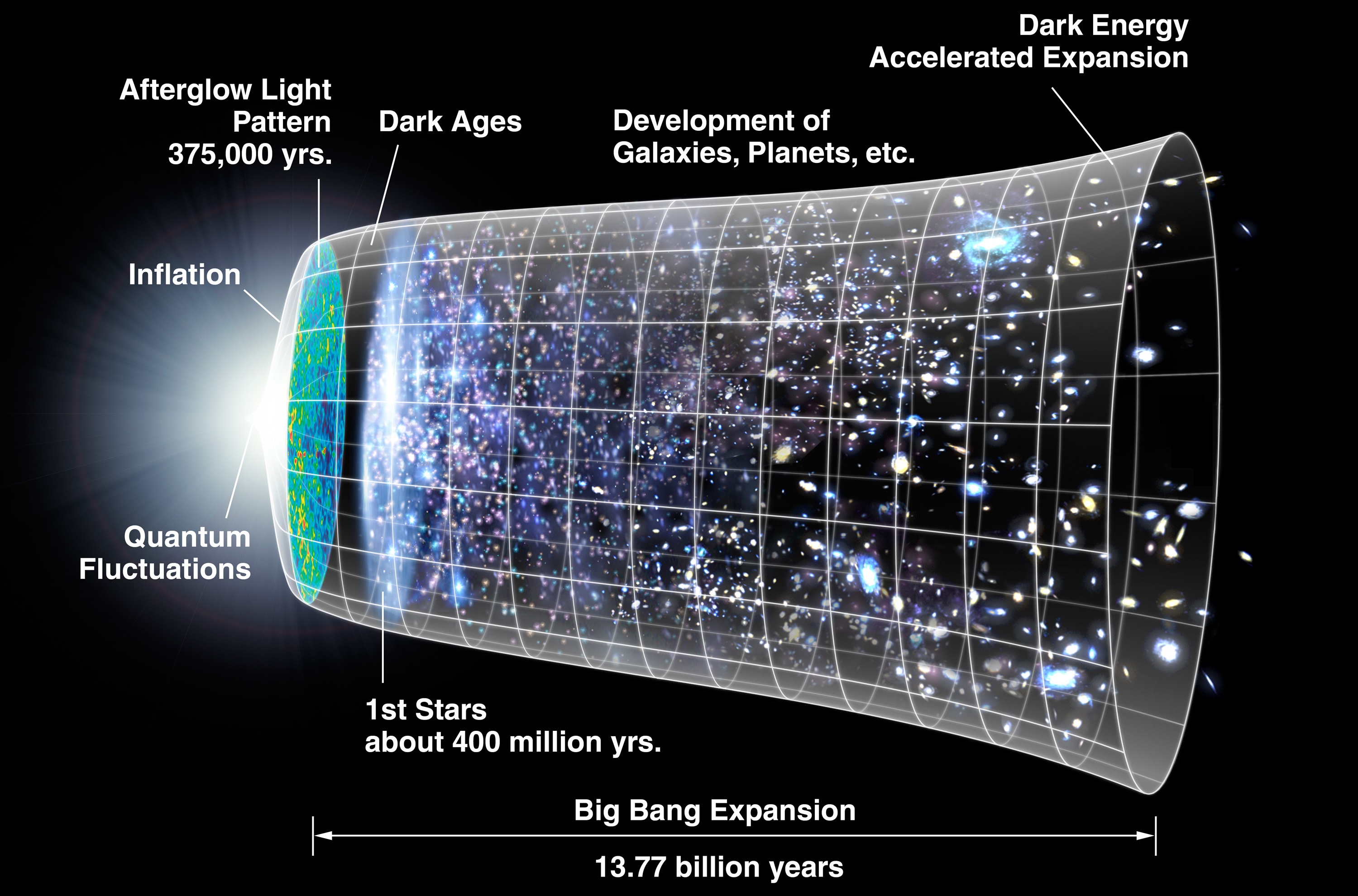}
	\caption{\textit{The evolutionary history of the universe, according to the Planck Team. Note that ``The age of the universe'' is taken to be a meaningful concept   by them, see} \cite{Planck2015},\cite{Planck2018}. From Wikipedia: \href{https://en.wikipedia.org/wiki/Chronology_of_the_universe}{Chronology of the Universe}}
	\label{fig:universehistory}
\end{figure}

\paragraph{The Passage of time in this context}
The process whereby the passage of time takes place at the micro level is irreversible collapse of the quantum wave function (Section \ref{sec:quantum}). The relation to the macro arrows of time is by a complex interweaving of bottom up emergence and top down constraints \cite{EBU}, as follows:
\begin{itemize}
	\item On the physics side, bottom up processes (described by coarse graining, effective field theories,  renormalisation group, crystalline bonding, chemical bonding, and so on) 
	lead to emergence of higher levels of structure and associated effective laws out of lower level structures and laws, with associated irrelevance of the microscopic details for what happens at larger scales;
	\item The higher levels 
	provide the context wherein lower level processes take place; this context exerts a top down effect on that dynamics both by setting global boundary conditions, and by exerting time dependent constraints on lower level processes \cite{Ell16};  
	\item In particular (Figure \ref{fig:universehistory}), a cosmological space-time structure emerges in a bottom up way through the time development associated with the local effects described by the Einstein Field Equations (which are 2nd order partial differential equations for the metric tensor \cite{HE}, hence represent local effects); the outcomes \cite{Dod03} \cite{PetUza13} \cite{Planck2015} \cite{Planck2018} are determined by initial conditions together with the changing equation of state of matter as time passes;
\item A global Direction of Time arises from the cosmological context  of an expanding and evolving universe; 
	\item This overall context, and particularly the associated decrease of temperature of interacting matter and radiation with time,  together with a Past Condition of low initial entropy, leads to local Arrows of Time (thermodynamic, electrodynamic, gravitational, wave,  quantum) coming into existence that are concordant with the global Direction of Time, thereby determining the direction of time for physical micro processes, including baryosynthesis, nucleosynthesis, and decoupling of matter and radiation;
	\item Finally emergence of complex systems occurs, such as accretion discs, stars, and planetary systems on the physical sciences side and  biomolecules, cells, organisms, ecosystems, and societies on the life sciences side, with the  thermodynamic, electrodynamic, and wave  arrows of time  determining  arrows of time in these emergent macro systems.
	\end{itemize} 

\noindent  As this happens in the context of the growing cosmological spacetime, time passes at each 
level that comes into being,
which implies it must also pass at the underlying microphysical level,
else the whole would not mesh coherently together.

\subsection{Time dependent constraints}\label{sec:_time_constraint}

Unitary evolution only occurs in very exceptional circumstances. What happens in fact in realistic contexts is that there is a coupling to the environment, which leads to time-dependent constraints and to dissipation, both of which affect dynamics on the microscale. We first focus on time dependent constraints $C(t)$. 
Instead of (\ref{eq:Hamiltonian}) one has
\begin{equation}\label{eq:Hamiltonian1}
V = V(\textbf{q},t)\,\,\Rightarrow\,\, H  = H(\textbf{p},\textbf{q},t).
\end{equation}
Consequently, there is no 
relation  (\ref{eq:unitary}); the real dynamics is 
\begin{equation}\label{eq:non_unitary}
\{\textbf{x}_0,\textbf{p}_0,C(t)\} \Rightarrow \{\textbf{x}(t),\textbf{p}(t)\}.
\end{equation}
In complex systems such time dependent constraints can express top-down effects whereby higher levels shape lower level physical outcomes \cite{Ellis_Kopel} \cite{Ellis_Drossel}.

\paragraph{Classical Mechanics} 
A simple example is a frictionless pendulum with  time dependent length $L(t)$. The equation of motion is  
\begin{equation}\label{eq:pendulum}
\frac{d^2\theta}{dt^2}(t)  + 2 \frac{\dot{L}(t)}{L(t)}\frac{d\theta}{dt}(t) + \frac{g}{L(t)} \sin\theta(t) = 0,
\end{equation}
and outcomes are determined by $\{\theta_0,\dot{\theta}_0,L(t)\}$ rather than $\{\theta_0,\dot{\theta}_0\}$.   
For details of the relation to the Hamiltonian, see the Appendix of \cite{Ellis_Kopel}. More generally, (\ref{eq:unitary}) will not be the case for any classical Hamiltonian system with a time dependent potential $V(\textbf{q},t): \,\,\partial V(\textbf{q},t)/\partial t \neq 0$.

\paragraph{Digital Computers} 
How this happens in practice in the case of digital computers is discussed in \cite{Ellis_Drossel}. A time dependent gate voltage $V(t)$ determined by the machine code alters the electric potential $V(\textbf{r}_i,t)$ in the Hamiltonian in a time dependent way; thus abstract algorithms can control physical outcomes. 

\paragraph{Biology} 
How this works  in the case of biology is discussed in \cite{Ellis_Kopel}. Membrane potentials  alter protein configurations and so change ionic distances, and hence alter the molecular Hamiltonian in a time dependent way \cite{Ellis_Kopel}. Similarly, messenger molecules alter biomolecule configurations \cite{Molecules} in response to higher level biological needs \cite{Hart99} conveyed by cell signaling networks \cite{Berridge}. 

\paragraph{Overall} To summarize so far: Non-unitary dynamics 
will occur whenever there is an explicitly time dependent Hamiltonian: the standard  uniqueness theorems \cite{Arnold}, \cite{Arnold1} for Hamiltonian systems with time independent potentials 
then don't hold: the initial state then does not uniquely fix the future evolution. And that is the case almost all the time in real physical and biological systems. Constant boundary conditions are an approximation that is only true on restricted timescales, even in the canonical case of planetary motion. 
The planets did not even exist if we go back far enough in the past.

\subsection{Ongoing coupling to the environment}\label{se: env_coupling}
In addition to time-dependent constraints, which set the context in which the time evolution of a system takes place, the environment influences a system also via dissipation and other ways of ongoing exchange of energy and possibly also matter. There is in fact no system that is fully isolated from these influences.  Let us look again at the three examples of the previous subsection:

\paragraph{Classical Mechanics} 
A classical mechanical system with friction does not undergo Hamiltonian time evolution, as its phase space volume shrinks. In the absence of time-dependent driving, friction extracts energy from a mechanical system, until it comes to a rest. The standard example is again a pendulum. In the absence of driving its amplitude decreases until the pendulum comes to a halt. With periodic driving, the phase space volume decreases also, but now the long-time behavior is given by limit cycles or strange attractors \cite{strogatz}. 

In the solar system, tidal forces are frictional forces that slow down the rotation of moons and planets and thereby accelerate their orbital motion. 

\paragraph{Digital Computers}
A digital computer needs a power supply in order to execute its tasks. Electrical currents, however, produce heat and thus dissipate energy. Furthermore, dissipation is essential for obtaining a reproducible response of a transistor to changes in the gate voltage. Otherwise the transistor could not forget its past, and its state would not be uniquely determined by the present gate voltage.

\paragraph{Biology}
Biological systems are prime examples of dissipative systems  that operate very far from thermodynamic equilibrium. They depend on their environment in such a strong way that they die if they are separated from it. There is an ongoing exchange of energy and molecules with the environment. This exchange is required to sustain the nonequilibrium structures that allow the processes mentioned in the previous subsection to happen. 

\subsection{Generically: Parts and the Whole}\label{sec:parts_whole}
\paragraph{Lack of Isolation of Parts} Taken together, the two previous subsections underline the fact that no level in the hierarchy of systems (apart from the Universe as a whole, see below) is completely isolated from the influence of higher levels.  This means that there is no unitary time evolution in practice: at best such a description is approximately valid for restricted systems and restricted times.
In the case of both digital computers and biology, the relevant lower level physics (interactions between electrons and ions) is conscripted to carry out higher level purposes \cite{Tannen},\cite{Hart99} in accord with the relevant higher level dynamics (e.g. execution of an algorithm, or beating of a heart) according to the timescale associated with that higher level dynamics (in the case of a computer, controlled by a clock \cite{Tannen};  in the case of the heart, controlled by a pacemaker \cite{Noble2002}). The time-dependent  constraint $C(t)$ in (\ref{eq:non_unitary}) sets the timescale for happenings at the electron level as determined by the higher level irreducible processes (the pacemaker, for example, functions as an integral system at its own emergent level; it cannot be reduced to lower level entities or dynamics). 

At neither level is the dynamics reversible. The pacemaker dissipates energy at the macro level due to irreversible metabolic processes at the molecular level, implemented by the underlying physics in this context. These top down processes cause the physics at the bottom levels to also proceed irreversibly.  
 However the speed at which things can happen at higher levels is constrained by the speed of the lower level processes. Hence there is a delicate interplay between timescales at the higher and lower levels \cite{EBU}. 

\paragraph{The Universe} What about the Universe as a whole? The view we take is that there is no macro Law of Dynamics for the Universe as a whole; rather the dynamics of the Universe is determined by coarse-graining the local effects of gravity everywhere, given the initial conditions. 
 This is what happens for example in the standard Friedmann-Lema\^{i}tre-Robertson-Walker (FLRW) models of cosmology, with their spatially isotropic and homogeneous geometry \cite{HE}. If you have any small volume of fluid $dV$ where the acceleration $a^\nu$, shear $\sigma_{ab}$, and vorticity $\omega_{ab}$ 
all vanish, then the Friedmann and Raychauduri equations  of  cosmology  will  apply  to  that  small  volume $dV$ (for  an  elegant covariant  proof,  see   \cite{Ell71}).   These  equations  apply  to  the  universe  on  a  large scale simply because they apply to each fluid element separately, and so arise by coarse graining the local dynamics. You can in this case determine these equations  by  applying  the  field  equations  to  the  large  scale  metric  (\ref{eq:FLRW}),  but that works because of the symmetry of the spacetime, not because that is the scale at which the equations apply. The Einstein Field Equations
\begin{equation}\label{eq:EFE}
R_{ab} - \frac{1}{2}R g_{ab} = \kappa T_{ab} + \Lambda g_{ab}
\end{equation}
are tested at the scale of the Solar System and only apply on larger scales by coarse-graining the small scale results, which generically adds a polarisation term ${\cal Q}_{ab}$ to (\ref{eq:EFE}), in parallel to how this happens in the case of electromagnetism.\footnote{For a parallel discussion of how this works in quantum theory, see \cite{Ell2012}.}
This back-reaction term that arises from the non-commutation of averaging and taking the field equations \cite{Ell84,Clarkson},  vanishes in the FLRW  case because of the high symmetry of those  spacetimes. 

Now  the  outcome depends on the behaviour of the matter tensor term $T_{ab}$ in the Einstein Field Equations, in turn determined by the equations of state for matter. And they   can be affected by the quantum uncertainty  discussed in Section \ref{sec:quantum}. 
Can that effect ever be important? Yes,  that is believed to have happened  in the inflationary epoch  in the very early universe \cite{Dod03},  \cite{PetUza13} when quantum perturbations provided the seeds for classical fluctuations at later times; and because this is a quantum effect,  those classical outcomes are in principle underterminable by the initial data at the start of inflation.  But then the key issue is that the resultant  inhomogeneities can act back on the background FLRW  cosmological model to change its expansion rate due to the back reaction effects mentioned above  \cite{Ell84}. The amplitude of this effect has been the subject of much debate, see \cite{Clarkson},  \cite{Back00}, \cite{Back0},  \cite{Back1}, \cite{Back2} and references therein. The effect is not large, but is seemingly non-zero, indeed it seems that it affects observational predictions in cosmology at the few percent level.

In summary: even though there is no higher level context for the Universe as a whole, random effects seeded by quantum fluctuations occur even at the cosmological scale. The statistical outcomes of structure formation are determined \cite{Planck2015} \cite{Planck2018}, but not the unique specific outcomes that actually happen.

\section{Evolving block universe and the Direction of Time}
\label{sec:EBU}
If time passes, as it does, there 
  must be a way to deal with this in a space-time picture, in a manner consistent with General Relativity theory. Indeed there is: it is an \textit{Evolving Block Universe}, or EBU (\cite{EBU1}, \cite{EBU}, \cite{EBU2}). The basic idea is that we consider a spacetime manifold with both future and past boundaries, where the future boundary keeps advancing as time progresses.

\subsection{The Evolving Block Universe}
\label{sec:Boundary}
The EBU is conceptually a spacetime manifold with a fixed past boundary and a moving future boundary. Technically it is a one parameter family of manifolds. 

\paragraph{Manifold with boundary} A 4-manifold ${\cal M}^+(t_0)$ with a future boundary $\{t = t_0\}$ is determined by a homeomorphism of a 4-dimensional topological space  to the half region of 4-dimensional Euclidean space $E^-:\{t\leq t_0\}$ \cite{Boundary}. Similarly for a manifold ${\cal M}_-(0)$ with a past boundary there is a homeomorphism  to the half region of 4-dimensional Euclidean space $E^+:\{t\geq 0 \}$, and for a manifold ${\cal M}(t_0,0)$  with both boundaries, a homeomorphism to the region of 4-dimensional Euclidean space $E^\star:\{t_0 \geq t\geq 0\}$.

\paragraph{The Evolving Block Universe} In the cosmological context we consider a 1-parameter family of manifolds $\{{\cal M}(t,0)\}$ with fixed past boundary boundary $\{{\cal B}:t=0\}$ and time dependent future boundary ${\cal P}(t)$. This is the evolving  block universe. Each manifold ${\cal M}(t_1,0)$ contains ${\cal M}(t_2,0)$ as a subset if $t_1 > t_2$.

\paragraph{Example} The simplest example is a Friedmann-Lema\^{i}tre-Robertson-Walker (FLRW) spacetime \cite{Ell71} \cite{HE}
\begin{equation}\label{eq:FLRW}
ds^2 = - dt^2 + a^2(t) d\sigma^2, \,\,\,0\leq  t\leq t_0, \,\,u^a = \delta^a_0,
\end{equation}  
where $d\sigma^2$ is a 3-space of constant curvature $k = \{+1\, \,or\,\, 0\, or \, -1\}$.
 The 3-surfaces $\{t = const\}$ are spatially homogeneous, the universe is spatially isotropic about every point, and the matter flow lines with tangent vector $u^a (u^a u_a = -1)$ are geodesic. The Einstein Field Equations (\ref{eq:EFE}) together with matter equations of state determine its evolution (\S\ref{sec:parts_whole}).   

\begin{figure}[h]
	\centering	 
	\includegraphics[width=0.7\linewidth]{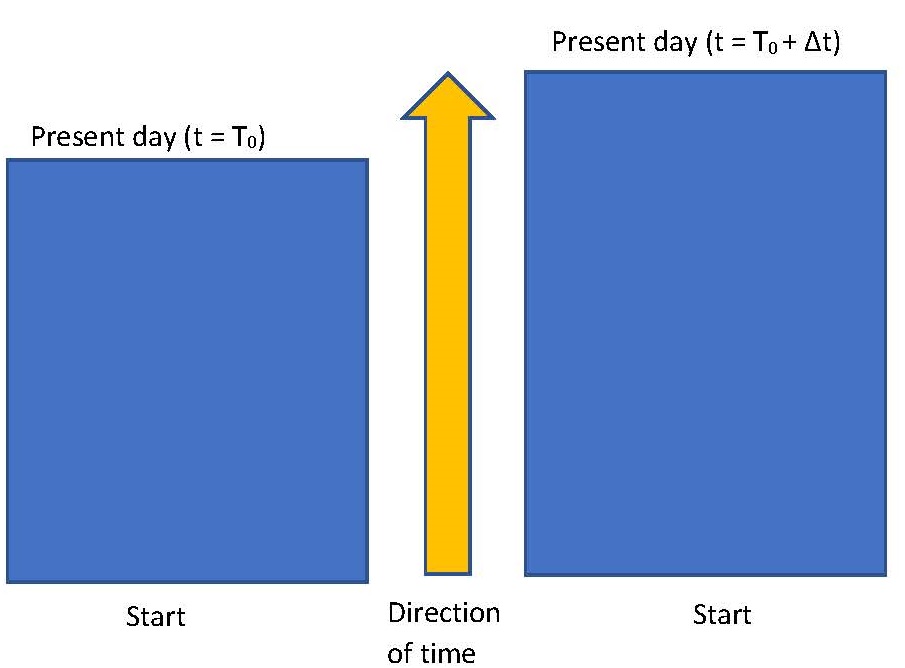}
	\caption{\textbf{The EBU and the direction of time.} \textit{The evolving block universe at time $T=t_0$ (left) and $t_1 = T_o + \Delta t$ (right).  Spacetime ${\cal M}(t,0)$ has two boundaries: a fixed one at the start of the universe $t = 0$ (bottom), and a moving one at the time $t$ (top). The } Direction of Time \textit{points from the fixed start to the ever changing present time. As $t$ increases, the manifold ${\cal M}(t,0)$ continually gets larger, with ${\cal M}(t,0)$ being a subset of ${\cal M}(t +\Delta t,0)$ }. }
	\label{fig:1}      
\end{figure}

The past boundary $t=0$ (the initial singularity) is fixed. The future boundary $t=t_0$ (the present time) continually extends to the future as time passes  (Figure \ref{fig:1}). At any specific time $t=t_0$, the spacetime manifold ${\cal M}(t_0,0)$ defined in this way exists from the start of the universe $(t = 0)$ to that time $t_0$. As time passes, the future boundary grows: $t_0 \rightarrow t_1 = t_0 + \Delta t, \, \Delta t >0. $ so the manifold is then ${\cal M}(t_1,0)$. 
The  Direction of Time is determined by this process: a manifold ${\cal M}(t_2,0)$ at a later time $t_2$ than the earlier time $t_1 < t_2$ contains the corresponding manifold ${\cal M}(t_1,0)$  as a subset. Hence one can order the manifolds by this inclusion mapping and determine the cosmological direction of time 
 for a family of manifolds ${\cal M}(t,0)$ through this ordering. 

\paragraph{The age of the universe} This spacetime view is implicit in standard cosmology such as the Planck satellite  analysis of CMB anisotropy data \cite{Planck2015} \cite{Planck2018} (Figure \ref{fig:universehistory}), even though they do not explicitly state this. 
The key point is that they give a figure $T_0$ for the age of the universe at the time that the measurement was made: $T_0 \simeq 13.87 \times 10^9$ years. This would not make sense unless the concept of the age of the universe (at the time of measurement) had a meaning (Figure \ref{fig:universehistory} and left hand figure \ref{fig:1}).
 If the experiment is repeated at a later time $T_1 = T_0 + \Delta T_0,\,\Delta T_0>0$, the age will have increased by $\Delta T_0$ and the EBU will have extended to the future by that amount (right hand figure \ref{fig:1}).

\paragraph{The past exists, the future does not}
The EBU  ${\cal M}(t_0,0)$ exists for all times $t: 0 < t < t_0$ because the  corresponding events lie in the causal past of the present $t = t_0$ and so are able to influence what happens then. For example, baryosynthesis, cosmological nucleosynthesis, first generation star formation, stellar nucleosynthesis, second generation star formation, planetary formation all took place prior to the present day $t = T_0$ (that is, 25  November 2019), and have all influenced conditions on this planet at this moment. These processes must as a matter of fact have taken place in the past (in an ontological sense: they actually happened), otherwise we would have uncaused features at the present day (the existence of baryons, carbon, and so on on Earth, and indeed the existence of the Earth itself). Accordingly the space time regions where they occurred must have existed then, else it would not have been possible for them to have happened. This is represented by the canonical Planck Figure \ref{fig:universehistory} and by the EBU (Figure  \ref{fig:1}). 

By contrast, the future region $t> t_0$ does not yet exist  because, due to the key feature of irreducible quantum uncertainty, what will happen then is not yet decided (Section \ref{sec:quantum}). Due to conservation laws there are restrictions on what can happen in the future, but which of those will occur is yet to be decided and, so it cannot influence the present. As this is true both for events in spacetime and for spacetime itself (Section \ref{sec:quanr_grav}), the future is not yet determined and so does not exist in an ontological sense. This view is opposed to those of  `Presentism' and `Eternalism' \cite{Rovell_present}.  

\paragraph{The far future} Ultimately, $t$ increases from $0$ to some maximum value $t_{max}$ such that spacetime  ${\cal M}(t_{max},0)$ is inextendible, either because the future boundary runs into a singularity after recollapse, or asymptotes to infinity; thus $t_{max}$ may be finite or infinite. Because of the presence of dark energy which is probably a cosmological constant \cite{Dod03} \cite{PetUza13} \cite{Planck2018}, and hence will cause  accelerated expansion for all future time, the latter is the most likely case.  
Standard conformal spacetime diagrams showing future infinity \cite{HE} \cite{Penrose_Road} represent this situation where${\cal M}(t,0)$ becomes ${\cal M}^\infty(0)$ in the far distant future: 
\begin{equation}\label{key}
{\cal M}^\infty(0) = \lim_{t\rightarrow\infty}{\cal M}(t,0).
\end{equation}
Thus the reason the present does not occur in these diagrams is because in them, time has fully run its course; everything that can happen has happened. Every present surface $\{t = t_0\}$ corresponding to events that have happened lie in the past of future infinity. By contrast, the concept of the present is explicitly built into the canonical NASA diagram (Figure \ref{fig:universehistory}) of the evolution of the universe, as presented in the Wikipedia article 
\href{https://en.wikipedia.org/wiki/Chronology_of_the_universe}{Chronology of the Cosmos}.

\subsection{Evolution along preferred timelike world lines}
However we need to be able to construct such an EBU for more general spacetimes, and particularly for the perturbed FLRW models that are our best models of the real universe \cite{Dod03} \cite{PetUza13}. A number of issues arise. 
 
 \paragraph{Relativity of simultaneity} What about the relativity of simultaneity in special relativity? This is often taken to be the death knell of such models, for it implies there can be no preferred spatial surfaces, such as the present $\{t = t_0\}$ in the EBU at each time $t_0$. A change of velocity will mean different instantaneous spatial surfaces, and hence one can't have such preferred surfaces.
 
 The response is that the spacetime structure of the universe is determined by General Relativity \cite{HE}, not Special Relativity, and all physically realistic spacetimes have preferred timelike world lines \cite{Ell71} and spatial surfaces, as in the case of FLRW spacetimes (\ref{eq:FLRW}). As in all real physics, the symmetry of the underlying theory is broken by physically occurring structures. In short, the real universe is not a de Sitter, anti-de Sitter, or Minkowskian spacetime \cite{HE}.  
 
 \paragraph{Evolution takes place along timelike world lines} 
 Actually surfaces of simultaneity determined by radar, as in Special Relativity, are irrelevant to dynamics because no influence travels faster than light. Constraint equations can hold on spacelike surfaces and are then conserved if they are initially true, but that is then a consequence of the dynamics.   
In fact dynamical evolution generally corresponds to influences occurring along timelike world lines; only plane wave modes cause effects on null geodesics. This does not occur significantly on cosmological scales.
 \begin{itemize}
 	\item 
 \textbf{Matter}: because matter has mass, its evolution takes place along timelike world lines; for example the 4-velocity of a perfect fluid. It has matter modes and sound wave modes \cite{Ehlers_Prasanna}. 
\item \textbf{Huyghen's principle}: From a differential equation viewpoint, the essential issue as regards all radiation is that Huyghen's principle for perturbations only holds under very restricted conditions: probably only for conformally flat or plane wave background spacetimes \cite{McLenaghan} \cite{McLenaghanI} \cite{McLenaghanII}. Thus it will only approximately hold in the real universe, and tails will occur, effectively meaning timelike propagation. Underlying this is the algebraic fact that except in the case of parallel null vectors, the sum of two null vectors or of a null vector and a timelike vector is a timelike vector, so when photons or gravitons interact with each other or other particles the outcome is timelike.
\item \textbf{Electromagnetic Radiation}: The geometric optics approximation with propagation along null geodesics \cite{Ell71} is corrected at next order by tails in all but very exceptional spacetimes as just indicated. During propagation in a plasma, the rays are timelike  (\cite{Breuer_EhlersI}, \cite{Breuer_EhlersII}). Furthermore black body radiation (a statistical sum of massless photons with a Planck frequency distribution)  has stress tensor 
\begin{equation}\label{eq:bbR}
T^{ab}= \sum_{(i),(j)} f(\textbf{k})k^a_{(i)}k^b_{(j)}, \,\,k^a_{(i)}k_a{}_{(i)}=0\,\,\Rightarrow T^a_{\,\,\,a}=0
\end{equation}
which in the case of an isotropic distribution is a perfect fluid with timelike eigenvector $u^a$  and equation of state $p = \rho/3$. 
\item \textbf{Gravitational Radiation:} In the case of gravitational radiation, generically tails will occur. If it interacts with matter, it is no longer freely propagating at the speed of light and  irreversible processes occur  \cite{Hawking66}.     
 \end{itemize}
Overall, in realistic cases dynamical influences in curved spacetime are along timelike curves, which are not geodesics except in special cases such as (\ref{eq:FLRW}).  Generically inhomogeneity will cause non-geodesic motion. 
 
\subsection{Surfaces of constant time} \label{sec:const_time}
So given this feature of effective timelike nature of causation, how does one determine surfaces of equal time?

The proposal made here \cite{EBU} \cite{EBU1} \cite{EBU2} is that in a generic perturbed FLRW universe model, surfaces of constant time $\{\tau = const\}$ are determined by  proper time 
\begin{equation}\label{eq:time}
\tau(v) = \int_0^v \sqrt{|g_{ij}(x^0(v),x^\nu_{\,\,0})dx^idx^j|}\, dv,\,\,\forall\,x^\nu_{\,\,0} 
\end{equation}
determined along a congruence of preferred timelike lines $x^a(v) = \{x^0(v),x^\nu_{\,\,0}\}$ from the start of the universe $\{v=0,x^\nu_{\,\,0}\}$ to the event $\{x^0(v),x^\nu_{\,\,0}\}$ starting at each spatial position $x^\nu_{\,\,0}$, where $v$ is an arbitrary curve parameter and $\{x^\nu\}$ are comoving coordinates. The prescription is to start at the initial singularity\footnote{In cases where there is no initial singularity, a surface of constant density $\rho$ that corresponds to a bounce if that happens; else in an emergent universe \cite{Emergent1} \cite{Emergent2}, an arbitrarily chosen surface of constant density that occurs way before inflation starts. } (by definition, $(\tau(0) = 0))$ and use integral (\ref{eq:time} along preferred timelike worldlines to determine the constant time surfaces $\{\tau = const\}$.  Thus the surfaces of constant proper time $\{\tau = const\}$ (determined by (\ref{eq:time})) are secondary to the timelike world lines $x^a(v)$. The time parameter $t$ in the previous sections will from now on be chosen to be proper time $\tau$ determined in this way.

In a FLRW universe (\ref{eq:FLRW}), these will be the standard surfaces of constant time $\{t=const\}$. However these surfaces of constant time $\tau$ will not be instantaneous in the radar sense if the universe is expanding or inhomogeneous, and generically may not even be spacelike. 
 
 It is crucial that proper time $\tau$ determined in this way is the physically meaningful time for all local physical processes \cite{HE}: indeed the whole purpose of the spacetime metric $g_{ij}(x^k)$ is precisely to determine the uniquely meaningful  physical  time along time like world lines via (\ref{eq:time})  (which then, because of the indefinite signature of the metric associated with Lorentz invariance, determines its causal structure). Thus proper time $\tau$ as in (\ref{eq:time}) is the time $t$ that occurs for local observers in Newton's laws of motion, Maxwell's equations, the Schr\"{o}dinger equation, the Dirac equation, and the 1+3 covariant form of the Einstein Field Equations \cite{Ell71}.  Indeed it is a key feature of physics that there is a single time function that occurs in all these contexts. 
 
\paragraph{The preferred timelike world lines}
To make this prescription geometrically and physically unique, one must defined a preferred family of timelike world lines $x^a(\tau)$. These are given \cite{EBU1} \cite{EBU2} by the family of fundamental world lines with tangent vector $u^a = dx^a/d\tau$ determined by   
\begin{equation}\label{eq:eigen}
T_{ab}u^a = \rho u_b \Leftrightarrow R_{ab}u^a = \mu u_b, \,\,u^a u_a = -1.
\end{equation}
 That is, $u^a$ is the timelike eigenvector of the matter stress tensor (the Landau frame), which by the Einstein Field Equations is also the timelike eigenvector of the Ricci tensor. Thus it breaks Lorentz symmetry: it is preferred in both physical and geometrical terms.
 
 What if spacetime is empty:
 \begin{equation}\label{eq:vac}
 \{T_{ab}= 0\} \Leftrightarrow \{R_{ab}=0\}, 
 \end{equation}
  so (\ref{eq:eigen}) is trivially true for all unit timelike vectors?  The response is that this is nowhere true in the real universe: \textit{inter alia} the cosmic background radiation pervades all space at all times since decoupling, and defines a unique timelike direction at every spacetime event.\footnote{It is true that closed buildings or boxes can exclude the CMB; however (a) they cannot occur on an astronomical scale, where in any event many other forms of radiation will generically occur and prevent a vacuum; (b) on a micro scale such a box cannot contain an exact vacuum for technological reasons, and it itself will move on a timelike worldline.  } However equation (\ref{eq:eigen}) will not determine a unique timelike eigen-direction for all mathematically possible matter tensors $T_{ab} \neq 0$; there are exceptional cases where this vector is not uniquely defined, but they do not correspond to physically realistic forms of matter. We  will take existence of a unique solution to (\ref{eq:eigen})  as a requirement, in effect a form of energy condition \cite{HE}, that must be satisfied if the matter tensor is to be considered physically realistic (as a specific example, the case of pure null radiation $T_{ab}= f(\textbf{k}) k_a k_b, \,\,k^a k_a = 0$ will not occur as the total energy-momentum tensor of a realistic spacetime, because of other matter and radiation that will be present).

\subsection{Spacetime Evolution}
Given that proper time is determined via the spacetime metric $g_{ij}(x^k)$, what determines the metric? It is determined by the matter present together with suitable initial conditions, via the Einstein Field Equations \cite{HE}.
\paragraph{Evolution equations}
Using the ADM formalism \cite{ADM} evolution equations for spacetime are determined after choice of a foliation of spacetime by surfaces $\{t = const \}$ and worldlines with tangent vector $u^a$, the relation to the normals $n^a$ being determined by the lapse vector $N^i(x^\alpha)$ and the relation of coordinate time to proper time along these world lines by the lapse function $N(x^\alpha)$ (details are given in \cite{ADM}). The application to the EBU is given in \cite{EBU2}. 

To correspond to the choice of vector (\ref{eq:eigen}), choose the 4-velocity to be a Ricci Eigenvector:
\begin{equation}\label{key}
T^\mu_0 = 0 \Rightarrow R^\mu = −2\pi^{μj}_
{|j} = 0, 
\end{equation}
which algebraically determines the shift vector $N^i(x^j)$, thereby solving the $(0,\nu)$ constraint equations. 
Determine the lapse function $N(x^i)$ by the condition (\ref{eq:time}) that the time parameter $t$ measures proper
time $\tau$ along the fundamental flow lines.
These conditions together uniquely determine the lapse and shift, as discussed in  \cite{EBU}. The evolution of matter is determined by the energy-momentum conservation equations (which are integrability conditions of the Field Equations)  together with the equations of state for the matter.  Then the usual ADM equations \cite{ADM} determine the spacetime metric from the given initial data. 

\paragraph{Equations of state and uniqueness} At a classical level, this is how spacetime and the Direction of Time emerge from the initial state of the universe. However what spacetime emerges depends on the equations of state relating the matter tensor components $\{\rho,p,q^a,\pi_{ab}\}$. As these equations of state can include quantum effects that are intrinsically indeterministic \cite{Ghirardi}, it might be that the specific future time development that actually occurs is in principle undetermined \cite{EBU1} \cite{EBU2}. Indeed this is the case in the real universe, because (assuming the inflationary picture is true \cite{Dod03} \cite{PetUza13}) quantum fluctuations during inflation led to classical fluctuations at the end of inflation by some as yet unknown process whereby collapse of the quantum wave function took place. This is an in principle stochastic event, but with well determined statistical outcomes leading to probabilistic  predictions for cosmological models \cite{Planck2015} \cite{Planck2018}. 

The implication is that the specific unique outcomes that actually occurred, such as the existence of our Galaxy, Sun, and Earth, is not uniquely specified by initial data in the very early universe at the start of inflation. However when quantum effects are not dominant and suitable classical equations of state are given, as at times after then end of inflation, outcomes will be unique \cite{HE}. However, this raises again the above-mentioned problem that deterministic equations require infinite precision, which is a mathematical and not a physical concept, see  \cite{Ellisetal_infinity} and \cite{gisin2018indeterminism}.

\paragraph{Synchronised `coming into being'?} The proposal here is in interesting counterpoint to Rovelli's proposal \cite{Rovell_present}  
	`\textit{`the fact that there is no preferred objective foliation of 4d spacetime into three dimensional `time instants' is not a denial of becoming: it is only a denial of a global synchronised becoming.''}

By contrast, we propose there is indeed a preferred objective foliation of a cosmological spacetime, but it is not synchronised in the sense of being instantaneous in the Special Relativity sense. Rather, coming into being takes place according to local physical processes based in proper time defined along preferred timelike world lines. No synchronisation process is involved.

\section{Arrows of time}\label{sec:arrows} 
\begin{figure}[h]
	\centering	 
	\includegraphics[width=0.7\linewidth]{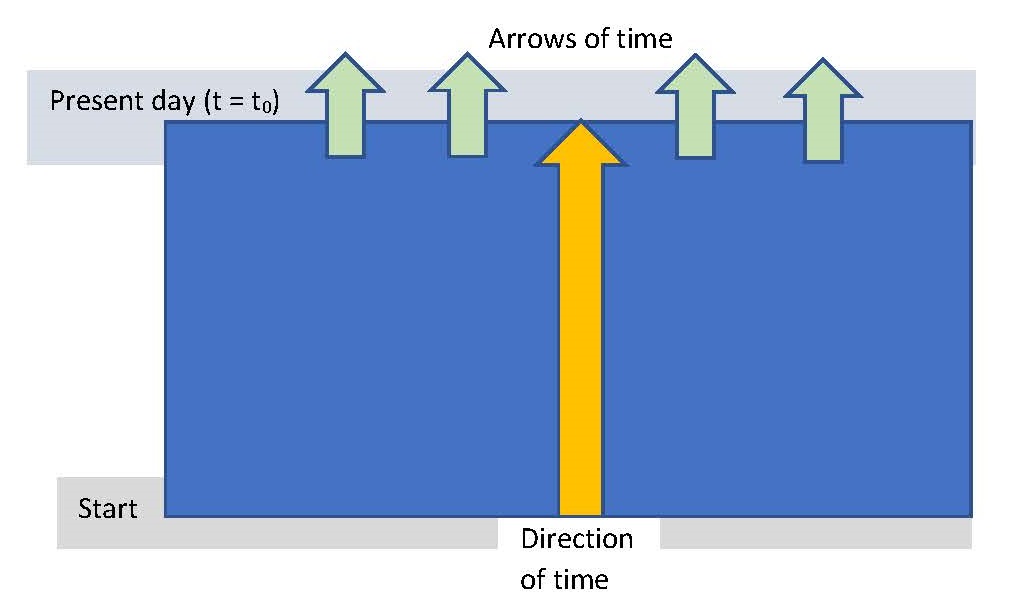}
	\caption{\textbf{Arrows of time}. \textit{Local arrows of time (thermodynamics, electrodynamic, wave, gravitational, and quantum) are determined by contextual effects so as to be aligned with the Direction of Time (Figure \ref{fig:1}) through various physical conditions discussed below.}}
	\label{fig:2}      
\end{figure}

Arrows of time are local physical effects which are  determined non-locally by the cosmological context of the EBU and the evolution of the universe as a whole (Section \ref{sec:Boundary}). This is indicated in Figure \ref{fig:2}.

The local arrows of time consist of the Thermodynamic Arrow of Time (\S \ref{sec:time_Thermod}) (entropy increases to the future), Electrodynamic Arrow of Time (\S\ref{sec:time_Electrodynamics}), Gravitational Arrow of Time (\S\ref{sec:time_grav}), other Wave Arrows of Time  (\ref{sec:arrow-waves}), Quantum Arrow of Time (\S \ref{sec:time_quantum}), and the  Biological Arrow of Time (\ref{sec:time_biol}), which are discussed in turn below. 

In each case time symmetric underlying basic equations lead to time asymmetric outcomes, in agreement with the Direction of Time,  due to the cosmological context.\footnote{We are  ignoring here the arrow of time associated with the Weak Force, which is weakly time asymmetric. This is an important issue to be tackled later. The justification for omitting it is that it does not directly impact the dynamics of every day life, but its role in the early universe (e.g. baryosynthesis) and in astrophysics needs consideration.} 

\subsection{Thermodynamic arrow of time}\label{sec:time_Thermod}
The Second Law of Thermodynamics  is a fundamental aspect of macrophysics, chemistry, and biology, as stated strongly by Arthur Eddington \cite{Eddington}. It relates to conversion of usable energy to unusable heat, usable materials to waste, diffusion of molecules, heat conduction, the heating effect of electric currents, and so on, and overall the  irreversible nature of naturally occurring  processes. It emerges from probabilities in phase space at the micro scale and their relation to probabilities at the macro scale \cite{Eddington} \cite{Penrose_Road} \cite{Penrose_Fashion}.  Let the  region ${\cal S}_{P,V,T}( \textbf{p}_i,\textbf{q}_j)$ of micro phase space  $\{\textbf{p}_i,\textbf{q}_j\}$ corresponding to coarse grained macro variables $(P,V,T)$ have volume $\{\cal V\}_{P,V,T}$. Hamiltonian evolution preserving a Liouville measure at the micro scale will be overwhelmingly likely to end up  in regions of phase space that have a very large volume $\{\cal V\}_{P,V,T}$ 
(Penrose \cite{Penrose_Fashion} \cite{Penrose_Road}). This assumption together with random initial conditions will lead to the Second Law (\ref{eq:2nd_Law}) with extremely high probability, even though it could in principle, in a quasi-equilibrium state, occasionally be violated by very large fluctuations.\footnote{
Tim Maudlin pointed out to us in a private communication that due to its statistical nature the second law of thermodynamics is not really a \textit{law}. This touches upon very interesting philosophical questions relating to the nature of physical laws in general and the second law of thermodynamics in particular that we will not pursue here. In fact, there is no general agreement on what precisely the second law of thermodynamics is \cite{uffink}.} However this derivation does not  in fact determine the thermodynamic arrow of time, because it works equally well in both directions of time. 

This situation changes when the deterministic time evolution is replaced by one with stochastic elements. This becomes necessary when infinite precision is abandoned since then the future states beyond a limited time horizon are not fixed by the initial state \cite{gisin2018indeterminism}. As soon as there is (objective) stochasticity, there are propensities for the different possibilities becoming actual. This presupposes a fundamental distinction between past and future, in harmony with the picture of an evolving block universe described above.

\paragraph{Loschmidt's Paradox}

The Second Law represents a key time asymmetry at the macro level \cite{Time_asymmetry} despite the time symmetry of the relevant underlying microphysical theories (classical mechanics, unitary quantum mechanics). How does the macrophysics know the direction of time, i.e. in what direction of time (\ref{eq:2nd_Law}) will hold, when the microphysics is time symmetric: that is, the symmetry \begin{equation}\label{eq:minus_t}
 T: \{t \rightarrow t' := -t\}
 \end{equation} leaves the underlying dynamical equations invariant. \textbf{\textit{Loschmidt's paradox}} is that if you take Boltzmann's derivation of the 2nd Law (\ref{eq:2nd_Law}) from kinetic theory, on using the substitution (\ref{eq:minus_t}),  precisely the same derivation also shows that $dS/dt' \geq 0$: that is, the Second Law also holds in the opposite direction of time. A preferred macro level arrow of time in which (\ref{eq:2nd_Law}) will hold cannot arise by coarse graining the micro level physics 
  because by (\ref{eq:minus_t}), both directions of time are equal as far as microphysics is concerned \cite{Penrose_Fashion}. 
  
More generally, time reversal symmetry in physics is discussed by Lamb and Roberts in \cite{Time_reverasl}. Consider a Hamiltonian $H(\textbf{q},\textbf{p})$ where $H(\textbf{q},\textbf{p})= H(\textbf{q},\textbf{-p})$, which is true with the usual kinetic term. Then the equations of motion (\ref{eq:Hamiltonian}) are  invariant under the transformation
\begin{equation}\label{key}
T: (\textbf{q},\textbf{p},t) \rightarrow (\textbf{q},-\textbf{p},-t)
\end{equation}
so if ${\cal S}^+(\textbf{q}(t),\textbf{p}(t))$ is a solution of the equations of motion with initial data  $(\textbf{q}(t_0),\textbf{p}(t_0))$, then  ${\cal S}^-(\textbf{q}(t),\textbf{p}(t)):= {\cal S}^+(\textbf{q}(-t),-\textbf{p}(-t))$ is a solution with initial data $(\textbf{q}(t_0),-\textbf{p}(t_0))$. That is, the system allows the identical dynamical trajectory but in the opposite direction of time. If coarse graining the first set of trajectories involves entropy increasing in the future, then coarse graining the second set will imply entropy increasing in the past. \footnote{In fact, the transformation \eqref{key} that is usually viewed as a time reversal transformation can also be interpreted differently: as Tim Maudlin pointed out to us, speaking of an evolution from an initial to a final state always defines a forward time direction, and in this sense time itself is never reversed, but the momenta and ensuing trajectories are. }

 The symmetry (\ref{eq:minus_t}) applies equally to the microphysical derivation of the Second Law by Boltzmann in the classical case, and by Weinberg in the quantum field theory case.\footnote{In the latter case, see \cite{Ell16}:281-282 for details.} Thus coarse graining does not by itself lead uniquely to a derivation of the second law of thermodynamics  with a unique direction of time associated with entropy increase \cite{Lebowitz}  \cite{Penrose_Road} \cite{Penrose_Fashion} \cite{Albert} \cite{Time_asymmetry}  \cite{Ell16}. It does not determine the thermodynamic arrow of time because by the underlying symmetry  (\ref{eq:minus_t}), these derivations equally deduce $dS/t \geq 0$ and $dS/dt' \geq 0$.

The solution involves two aspects: a past condition, that is, special initial conditions at the start of the universe, and a reconsideration of macro-micro relations, when time dependent constraints $C(t)$ 
negate the above argument: (\ref{eq:non_unitary}) holds rather than (\ref{eq:unitary}). 
Furthermore, irreversible stochastic events at the microlevel destroy the deterministic assumptions underlying the argument, as we will discuss below in Section \ref{sec:quantum}.

\paragraph{A Past Condition} The basic solution as regards physics in the early universe is a cosmological hypothesis:  boundary conditions are temporally asymmetric. Specifically, entropy was much lower in the very distant past \cite{Time_asymmetry} \cite{Penrose_Road} \cite{Penrose_Fashion} \cite{Rovell_entropy}, which is of course necessary in order that entropy can grow: if you start off in a state of maximum entropy, then entropy can't increase. This is  David Albert's \begin{quote}
	\textit{\textbf{Past Hypothesis}} \cite{Albert} \cite{Time_asymmetry}: ``\textit{We make the cosmological posit that the universe began in an extremely tiny section of its available phase space}.''
\end{quote}  
This provides the basis on which entropy can increase in the future direction of time during irreversible processes such as nucleosynthesis and decoupling of matter and radiation, and at later times as structure formation takes place. 
How these special initial conditions occurred is however a matter of contention, with some claiming inflation will solve it, and Penrose claiming that, because of the gravitational entropy associated with black holes, this is not the case \cite{Penrose_Fashion}. This comment may well be valid, 
but there seem to be  unsolved issues
with his proposed solution of a Conformal Cyclic Cosmology. 
In any case, however it happened, and however it is related to the definition of gravitational entropy \cite{Earman}, it is clear that such a condition must have occurred and provided the basis for the thermodynamic arrow of time (\ref{eq:2nd_Law}) in accord with the cosmological direction of time (Figure \ref{fig:2}).

\paragraph{Macro-micro issues} The second key point is  the relation of all this to macro-micro relations, where time dependent constraints occur (Section \ref{sec:micro_macro}), leading to an associated  direction of time downward cascade between scales (\cite{EBU}, \cite{Ell16}:280-284). For example the expansion of the universe (\ref{eq:FLRW}) in accord with the Direction of Time (\S\ref{sec:EBU}) sets an  arrow of time for nucleosynthesis and decoupling of matter and radiation by causing the cosmological  density $\rho(t)$ and temperature $T(t)$ to decrease with time:
\begin{equation}\label{eq:DTbydt}
\{da(t)/dt > 0\} \Rightarrow \{d\rho(t)/dt <0,\,\,dT(t)/dt < 0\},
\end{equation}  
where $da(t)/dt$ is determined from the cosmologically averaged energy density $\rho(t)$  and $d\rho(t)/dt \Rightarrow dT(t)/dt$ is determined by the cosmological equation of state $p=p(\rho,t)$.  
This overrides fluctuation effects and leads to the occurrence of 
irreversible processes with an associated time arrow,\footnote{The equation for the rate of change of the matter specific entropy is (3.13) in \cite{Ell71}.}  including the electroweak phase transition, quark-hadron phase transition, cosmological  nucleosynthesis, decoupling of matter and radiation, and formation of stars and galaxies as the temperature of the universe decreases \cite{Dod03} \cite{PetUza13}, with concomitant entropy increase at those times.
Gravity plays a central role here,
as it 
leads to the formation of stars and the occurrence of nuclear reactions in stars. These reactions, which transform lighter particles to heavier ones and concomitant  radiation, happen so slowly that the universe is caught in nonequilibrium 'hangups' for billions of years, even though this transformation of gravitational energy into other energy forms increases entropy and is favored thermodynamically   
\cite{dyson1971energy}. 
\footnote{Various authors suggest to introduce an entropy of the gravitational field in order to follow the way entropy changes during these processes \cite{Clifton}, 
\cite{Penrose_Fashion}. We will not pursue  that issue here.} 

Channels for various types of reversible processes (particle physics, nuclear physics, atomic physics, molecular) close off as the universe evolves due to the cosmic temperature  $T(t)$ falling below the threshold for each of these processes, allowing the building up of metastable thermodynamic states   as the temperature drops even further \cite{Rovell_entropy}. This furthermore leads to the existence of a dark night sky  
at later times once stars have formed,  
providing the heat sink into which waste heat is radiated by stars, the Sun, and the biosphere on Earth in the future direction of time (linking this to the electrodynamic arrow of time).  

This downward cascade introduces the thermodynamic arrow of time into physical chemistry and biochemistry processes at the microlevel through the crucial  role of heat baths in local physical processes (left hand side of Table \ref{Table2}). It then cascades up through emergent biochemical, developmental,  and physiological processes such as developmental programs, metabolic processes, and cell  signalling processes to introduce a concordant arrow of time in plant and animal physiology and the brain \cite{EBU} (right hand side of Table \ref{Table2}).

\subsection{Electrodynamic arrow of ime}\label{sec:time_Electrodynamics}
The electrodynamic arrow of time is the statement that electromagnetic waves are received after they are sent rather than before.  Maxwell's equations are however time symmetric, so do not lead to this result \cite{Wheeler_Feynman}: they lead to wave equations for  $\textbf{E}$ and $\textbf{H}$ in a source free region that are time symmetric, with propagation speed the speed of light $c$. Advanced and retarded potentials and associated Green's functions are equally solutions to these  equations, because of the symmetry (\ref{eq:minus_t}). They potentially allow electromagnetic   waves to travel into the past as well as the future at the speed of light  \cite{Ellis_SCiama}  \cite{Ellis2002}, and so to convey energy and momentum in both directions of time. 
 
The EBU provides a simple solution to this problem: one cannot determine propagation of electromagnetic waves using advanced potentials because the spacetime region over which the associated integral would have to be taken does not yet exist (\S\ref{sec:Boundary}). Only the retarded Green's function makes sense in this context, and this ensures that EM waves propagate to the future and not the past, in accordance with the cosmological Direction of Time. This is why transfer of heat and momentum  by radiation can only take place in the future direction of time.

To be more precise, the retarded Green's functions must be supplemented with appropriate boundary conditions since any solution of Maxwell's equations can be obtained using any Green's function if it is combined with suitable boundary conditions. The boundary condition to be used with the retarded Green's functions is the Sommerfeld radiation condition, which states that 'the energy which is radiated from the sources must scatter to infinity; no energy may be radiated from infinity into ... the field' \cite{Sommerfeld}. This condition is  in agreement with the second law of thermodynamics: Taking the retarded solutions with the Sommerfeld condition means that all sources of electrodynamic fields are localized sources that lie in the past \cite{weinstein}. From these sources, they propagate, and when they hit a macroscopic object they are absorbed and transformed into undirected thermal motion. Never do different walls 'conspire' to coherently emit radiation at the expense of thermal energy. This would be the time-reversed process that violates the second law of thermodynamics.

\subsection{Gravitational arrow of time}\label{sec:time_grav}
The gravitational arrow of time is associated with gravitational waves which travel at the speed of light  
in a time asymmetric way: the sources of the gravitational waves detected by LIGO are in the past, not the future, and they are localized in space, similar to sources of electromagnetic radiation. 
 However Einstein's gravitational field equations, and the resultant wave equations for the electric and magnetic parts of the Weyl tensor in a vacuum spacetime   \cite{Hawking66} have no preferred direction of time; by (\ref{eq:minus_t}), gravitational wave propagation could equally in principle be in the other direction of time \cite{Ellis_SCiama}.
   
The answer is as in the electromagnetic case: only retarded potentials make sense in the EBU context, because the future does not yet exist. The gravitational wave arrow of time is therefore necessarily in accordance with the cosmological direction of time. Note that this occurs also in tidal forces: it's just that the speed of light is so slow compared with tidal interactions that this makes no practical difference.\\

\subsection{Other wave effect  arrows of time}\label{sec:arrow-waves}

Waves are ubiquitous in the physical world \cite{waves}. Again the issue arises, Where does the associated arrow of time come from, given that the source-free wave equation is time symmetric? 

Sound is heard after it is emitted,  
The wave operator $\square$ of the wave equation for sound with speed of sound $c_s$   \begin{equation}\label{eq:wave}\square u := \frac{1}{c_s^2}\frac{\partial^2 u}{\partial t^2} -  \Delta^2 u
\end{equation}
 is time symmetric, so solutions to $\square u = 0$ should occur equally for both directions of time: that is, with the effect felt either before or after the signal was emitted.
 
The EBU story works again in this case too, as in the electromagnetic case: there is no future spacetime region from which influences can propagate from the future to affect the present. 
In terms of boundary conditions, this means again that all sound waves are due to localized sources in the past, and that absorption and dissipation processes eventually turn the energy of sound waves into thermal energy.

The same is true for all other waves in elastic media, such as earthquakes or water waves, which are emitted from a localized source in the temporal past, and thus arrive after their cause. 
Such waves are in general dissipative, as represented by the addition of a term proportional to $\partial u/\partial t$ in the wave equation. This term  is related to energy loss and the Second Law. 

\subsection{Quantum arrow of time}\label{sec:time_quantum}

The indefinite future changes to a definite outcome
when wave function collapse 
takes place \cite{Ghirardi} \cite{Ell2012}, as discussed in Section \ref{sec:quantum}. This happens in a contextual way, mediated by interactions with heat baths, which link the quantum arrow of time to the thermodynamic arrow of time \cite{CWC}.

The process is irreversible:  there is a loss of information about the initial state     during wave function collapse \cite{Ell2012}. This irreversibility  is evident for example in  
biological processes  such as photon detection in the eye by rhodopsin and  photosynthesis in plants by chlorophyll, both based in the release of electrons in response to  incoming photons.  
A specific electron gets released at a specific time and place when this occurs, with associated energy transfer to the eye or the plant; the process does not take place in the reverse direction of time because of the extreme conspiracy that would be required in terms of time reversed particle motions to release a photon.

\subsection{Biological arrow of time}\label{sec:time_biol}
The biological arrow of time follows primarily from the thermodynamic arrow of time, as for example in diffusion of molecules and conversion of energy to heat. But the quantum arrow also plays a role, for instance when photons are absorbed or transitions to a different state occur in molecules.

Developmental programs and processes such as duplication and reading of genes and cell signalling and metabolic processes \cite{CamRee05}  all have consequent arrows of time, leading to processes such as mental remembering and forgetting at the macro level. This is the upward cascade of arrows of time from micro to emergent levels \cite{EBU}  in agreement with the cosmological Direction of Time. \\

\noindent Overall, what we have is emergence of quite new properties from the underlying physics \cite{Ell16}, with broken symmetries being a key feature allowing this to occur \cite{Anderson}. The context of the EBU and its Direction of Time (Figure \ref{fig:1})) breaks the symmetry (\ref{eq:minus_t}), and (given the Past Hypothesis) leads to local arrows of time that are in concordance with the cosmological Direction of Time (Figure \ref{fig:2}). The cosmological context acts down to affect local physical happenings in crucial ways \cite{Ellis2002} \cite{Ell16}. The expansion of the universe results in  the dark night sky that provides the sink for waste heat from stars, the Sun, and our biosphere, that is radiated away in the infrared in the future direction of time.

\section{Quantum Issues}\label{sec:quantum}
Quantum theory has two parts (\cite{Penrose_Road}:527-533): unitary wavefunction evolution $U$, plus wave function reduction $R$  (\S\ref{sec:collapse}). The latter leads to definite physical outcomes, and so is associated with the passage of time (\S\ref{sec:qu_time}). One must consider however whether quantum physics is relevant to the passage of time (\S \ref{sec:quantum_relevance}), and if so, whether it has an intrinsic direction of time built into it (\ref{sec:quantum_time_direction}).  

\subsection{Collapse of the wave function}\label{sec:collapse}
The unitary part $U$ of the evolution of the quantum wave function,
\begin{equation}\label{eq:Schroedinger}
U:\,\,H |\Psi\rangle = i \hbar\frac{\partial}{\partial t}|\Psi\rangle ,
\end{equation}
 is described by the Schr\"{o}dinger equation in the non-relativistic case and 
 relativistic wave equations or propagators of quantum field theory 
 in the relativistic case \cite{Penrose_Road} \cite{QFT}. However as the very purpose of the wave function is to determine probabilities of classical outcomes, it  means nothing physical unless wave function reduction $R$ occurs \cite{Penrose_Road} \cite{Ell2012}. 

A simplified description of the process $R$ is as follows: When an event $R$ happens, a wave function $|\Psi\rangle$ that is a superposition of orthonormal  eigenstates  $|u_n\rangle$ of some operator is projected to a specific eigenstate $N$ of that operator:
 \begin{equation}\label{eq:collapse}
R:\,\,|\Psi\rangle(t_0) = \Sigma_n c_n |u_n\rangle \,\rightarrow |\Psi\rangle(t_1) = \alpha_N |u_N\rangle. 
\end{equation}
where $\alpha_N$ is the eigenvalue for that eigenvector. In reality, this process is far more complex, as discussed in \cite{CWC}. 
There is irreducible uncertainty in this irreversible process:  the specific outcome $N$ that occurs is not determined uniquely by the initial state $|\Psi\rangle(t_0)$ \cite{Ghirardi}. However the statistics of the outcomes is reliably determined by the Bohr rule:  the probability $p_N$ of the specific outcome $|u_N\rangle$ occurring is given by  
\begin{equation}\label{eq:Born}
 p_N = |c_N|^2.
 \end{equation}

\begin{figure}[h]
	\centering
	\includegraphics[width=0.6\linewidth]{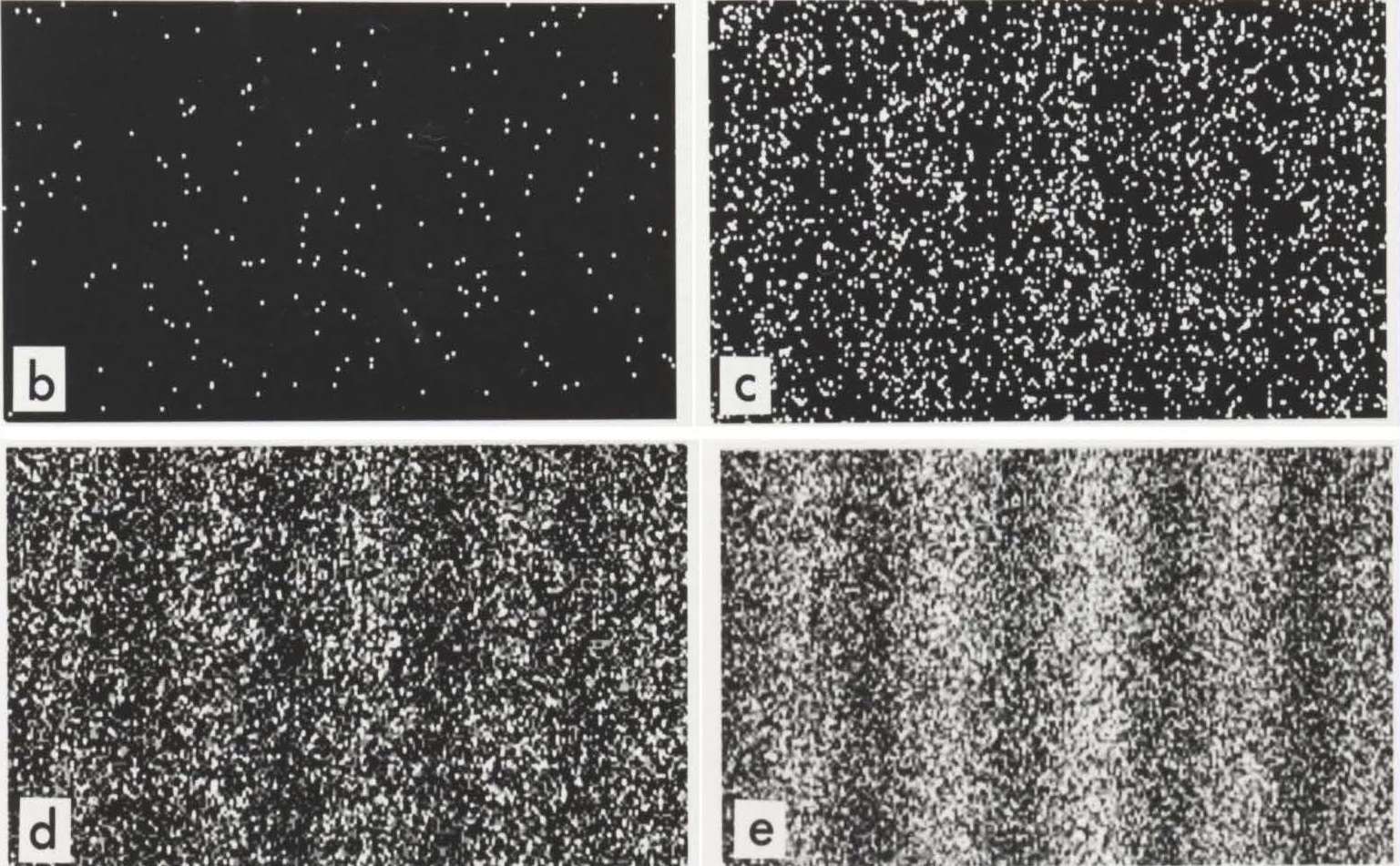}
	\caption{\textbf{Quantum uncertainty} \textit{Double slit experiment performed by Dr. Tonomura showing the build up of an interference pattern of single electrons. The numbers of electrons are, (b) 200, (c) 6000, (d) 40,000, and (e) 140,000.}}
	\label{fig:quantum_uncertain}
\end{figure}

This is shown in Figure \ref{fig:quantum_uncertain}, where the randomness of the individual events on the screen is clear (there is no known way to predict uniquely where the next one will occur) but the reliability of the statistical outcomes is apparent as the number of electrons that has gone through the slit increases and the classical wave interference pattern emerges. The projection process (\ref{eq:collapse}) happens in laboratory experiments such as the 2-slit experiment, but also occurs all the time when physical interactions take place such as   nucleosynthesis in the early universe, when a photon is registered by a CCD,  and when a photon hits a chlorophyll molecule in a leaf and hence releases an electron that starts a cascade of biochemical reactions during \href{https://en.wikipedia.org/wiki/Photosynthesis}{photosynthesis} in plants. Thus non-unitary events take place, that cannot be described by (\ref{eq:unitary}). The Copenhagen interpretation accepts this  description of quantum events \cite{Isham} \cite{Ell2012} \cite{Ghirardi}.

Now there are various alternative theories on the market to explain this experimental outcome, including hidden variable theories such as de Broglie-Bohm pilot wave theory (there is an inaccessible hidden variable underlying these statistical outcomes) and the Everett many worlds theory (the wave function of the universe splits into separate non-interacting  parts each time a measurement event takes place).  The latter is particularly extravagant because in many versions it is supposed to lead to the observer splitting into multiple observers each time a measurement takes place \cite{Isham}. However in solid physics terms both theories have no cash value, as both postulate the existence of entities that cannot even in principle be observed experimentally.  Furthermore, they cannot do with mere unitary time evolution, as both need to account for the observed stochasticity in one way or the other: The many worlds theory does it through the random choice of the trajectory that our consciousness takes at each branching event, and the pilot wave theory includes stochasticity through the infinitely many random digits that fix the initial state of the particle.   

We rather support   \textit{Contextual Wavefunction Collapse} \cite{CWC} where 
a nonunitary, stochastic  process (\ref{eq:collapse}) indeed takes place and obeys the Born rule (\ref{eq:Born}), 
but the way this happens is determined by the  local physical context. Indeed this is obviously the case:  specific apparatus may measure energy or polarisation, and outcomes depend on this choice; in the latter case the  direction of polarization measured can be chosen at will, again altering outcomes \cite{Susskind}. In effect this is a specific form of the Copenhagen interpretation where the macro apparatus is classical rather than quantum. The reason for this is limitations of the domain of validity of any particular wave function $|\Psi\rangle$ \cite{Ell2012}, and in particular the fact that a heat bath cannot be described by a many particle wave function \cite{Drossel_ten}. Any real macro apparatus involves heat baths and so is a classical entity, even though it emerges from a structure that has quantum properties on the microscopic scale.

\subsection{The passage of time}\label{sec:qu_time}
The proposal now is that in the semi-classical case, that is when we have quantum processes taking place on a classical spacetime background,  the passage of time takes place through the process $R$ described by equation (\ref{eq:collapse}) where the indefinite future changes to the definite past due to 
collapse events. 
As emphasized above, this process is not restricted to ``experiments'' carried out in a laboratory: it takes place all the time everywhere in the real world as physical, chemical, and biochemical interactions take place. The way it takes place is determined in each case by the local physical context, which breaks Lorentz symmetry: it is always associated with a preferred timelike direction (for example, the rest frame of a laboratory). 
Heath baths and their specific properties \cite{Drossel_ten} play a key role in this collapse process \cite{Ellis_Drossel},  providing the link to the thermodynamic arrow of time and so to the expansion of the universe. 

Furthermore, while this happens, quantum uncertainty at the micro level can get amplified to give unpredictable outcomes at the macro level. An example is that cosmic rays sometimes cause errors in transistor and hence computer  functioning  \cite{Cosmic_rays} in an inherently  unpredictable way, because emission of a cosmic ray is a quantum event that is unpredictable in principle \cite{Ghirardi}. 

\subsection{Relevance of quantum mechanics  to the passage of time?} \label{sec:quantum_relevance}
In response to the above, Carlo Rovelli (private communication) says,  \begin{quote}
	``\textit{I am very skeptical of the claim that quantum mechanics has anything to do with the arrow of time: if the macroscopic universe was classical, well described by classical GR, classical electromagnetism, and classical statistical mechanics hence thermodynamics, it would be nearly the same, with the same arrow of time. Therefore it seems to me that quantum mechanics is not relevant at all here...   The phenomenology of classical hard spheres spreading around in a box if they were initially concentrated is as irreversible as anything else, and heat plays no role... all the phenomena where you disregard heat are time reverse invariant.  For instance it is impossible to detect if the film of the motion of a pendulum, or of a planetary system or a few bouncing balls, is played forward or backward.   But as soon as there is a bit of dissipation, the pendulum slows down, the moon moves tides and moves further away, the balls slow down, in all these cases there is heat production and we can detect if the movie is played forward of backward.  Hence heat is part of the story of irreversibility.   More precisely: irreversibility comes when we have a macroscopic/microscopic distinction and the initial entropy is low}''.
\end{quote}
Now we agree with all except the claim of quantum irrelevance. Yes classical physics leads to the arrow of time; but after all it is emergent from quantum physics in a macro/micro context. The whole edifice is incomplete unless that quantum  foundation is included in the overall picture (surely we do not want an inconsistency between the quantum and macro levels?). That is what is achieved by the Contextual Wavefunction Collapse (CWC) view \cite{Ellis_Drossel}. The way quantum theory  relates to dissipative processes (and hence heat)  as mentioned by Rovelli is discussed in the specific case of a transistor in \cite{Ellis_Drossel}.

 In fact, no derivation of the second law from classical mechanics alone can do
without the concept of probabilities. Special trajectories that lead to
highly ordered end states are not forbidden by the laws of classical
mechanics, and within any given finite precision the initial state is
compatible with such a special evolution. We need therefore an
additional rule which tells us which types of trajectories are actually
taken. And here one cannot do without arguing with (objective,
intrinsic) probabilities. Why should the system care about our
subjective probabilities based on ignorance? If our guess was not
objectively correct, our predictions based on it would be false. But
if we need to introduce probabilities, we need a basis for them, and this
is found in quantum mechanical 'collapse' events.

Another (related) way
of arguing is that all the 'derivations' of the second law from
classical mechanics rely on limited precision of phase space points,
otherwise one would not have an ensemble of initial states from which to
single our a random one by the subsequent time evolution. Again, to
justify finite-size phase space points one ultimately needs  quantum
mechanics. 
These issues are discussed in depth in \cite{Drossel_2014}.

\subsection{A quantum direction of time?}\label{sec:quantum_time_direction}
The precisely opposite view is given by Donoghue and Meneze in \cite{Donoghue}: ``\textit{Hidden in our conventions for
	quantization is a connection to the definition of an arrow of causality, i.e., what is the past and what is the
	future.}'' 

This conclusion is not based on quantum wave equations alone, which describe a time-symmetric world, but also on propagators, which are an essential part of quantum field theory. By including a term $\pm i\epsilon$ in the denominators of propagators, one does the same thing as in classical electrodynamics when the Greens functions are calculated. The sign of the $i\epsilon$ term determines a direction of causality, with one option representing `retarded' and the other one `advanced' propagators. In one case the localized `source' of the propagator is in the past, in the other case in the future. A temporal direction of causality is explicitly included in the formalism of quantum field theory by choosing the appropriate propagators. The authors emphasize that the direction of causality is that of the thermodynamic arrow of time `because the increase of entropy occurs in
the direction that causal processes occur'. 

This is very similar to what was described further above in the context of classical waves: they emerge from a local source and propagate into the future and lead to an increase in entropy when they interact with a medium and deposit energy in it. 
In our previous paper \cite{CWC} we have also emphasized  that the creation and absorption events that prepare or measure particles represent nonunitary, entropy increasing events. 
The combination of unitary time evolution and $R$ events is thus built into the formalism of quantum field theory.\footnote{The authors comment furthermore on higher-order theories that can include both types of propagators and thus both causal directions. In this situation, there is causal uncertainty on short timescales.}
However, what is missing in \cite{Donoghue} and in any discussion that relates temporal arrows of wave propagation with the thermodynamic arrow is the connection to the cosmological arrow, which is responsible for aligning all local arrows of time with the global arrow imposed by the expansion of the universe. 

\section{Quantum gravity issues}\label{sec:quanr_grav}
Finally of course, space time itself should be emergent as time progresses. Figure \ref{fig:1} should refer not just to event in spacetime, but to spacetime itself.

Now of course this means we need a theory of quantum gravity, and we do not at present have a well-defined and consistent theory of quantum gravity \cite{MuruganFoundation}. In lieu of such, much has been made of the Wheeler-de Witt equation as describing the quantum wave function of the universe, and its unitary nature has been taken as a key argument as to why time does not pass \cite{Barbour}. However there is no evidence whatever that this equation is a good approximation to the real theory of quantum gravity in appropriate circumstances, and its proponents resort to the Everett (many worlds) view in order to try to make its use viable, despite there not being a shred of evidence that that equation describes any real physical system.

In contrast, one can suggest  that a viable quantum gravity theory should be based in discreteness of spacetime structure at the outset  \cite{Discrete}, and there are various options here \cite{Min_length}. One that accords with the EBU idea is the proposal of spin foams \cite{SpinFoam} \cite{SpinFoam1} which are discrete spacetimes that grow in the EBU sense, and so are compatible with what we are proposing.  

Thus while this (like all other quantum gravity proposals \cite{MuruganFoundation}) is still not a fully developed viable theory, it suggests directions to go that are  compatible with the proposals in this chapter. However a real challenge for any quantum gravity theory is to deal with the wave function collapse issue. Obviously we believe this should be tackled along the lines of Contextual Wavefunction Collapse \cite{Ellis_Drossel}. It could well be that Penrose' suggestion that gravity will be the relevant contextual feature associated with wave function collapse \cite{Penrose_Fashion} might turn out to be correct in this context.

\section{Conclusion}\label{sec:conclude}
This chapter has provided an integrated view of the passage of time at different levels of structure and function, with emergence of the Direction of Time at the macro (cosmological) scale (Figure \ref{fig:1}),  and concordant Arrows of Time at the local scale (Figure \ref{fig:2}). The associated spacetime picture is that of an Evolving Block Universe, which is what is in fact represented by the Planck team's canonical picture of cosmological evolution (Figure \ref{fig:universehistory}). Issues remain of course, such as the time asymmetry associated with the weak force. We believe the framework presented here is a good basis on which to investgate that issue. \\

The EBU proposal made here has been based in a particular proposal of how to determine preferred timelike world lines and derivative surfaces of constant time. For those who find that proposal inadequate, the challenge is to find some alternative proposal that does justice to the fact that time does indeed pass at the macro scale, and hence it must be represented by something like the EBU structure presented here.   To deny that time passes is to close one's mind to a vast body of evidence in a spectacular way. If we take that evidence seriously, either the view presented here, or some variant of it, must be the case.

\section*{Acknowledgements} We thank Carlo Rovelli,   John O'Donoghue, John Miller, and Tim Maudlin  for useful comments, and  Reinhard Stock for proposals that have substantially improved the text.

\end{document}